\documentclass[11pt,twoside]{article}
\usepackage{cal05}

\contact{Gerhardt R. Meurer}
\email{meurer@pha.jhu.edu}

\newcommand{\fion}[2]{\mbox{\rm [{#1}\thinspace{\footnotesize {#2}}]}} %Forbidden ion
\newcommand{\Hline}[1]{\mbox{H{\footnotesize {#1}}}}
\newcommand{\Halpha}{\Hline{\mbox{$\alpha$}}}

\begin{document}

\title{Selection and Characterization of Interesting Grism Spectra}

\author{Gerhardt R.\ Meurer}
\affil{Department of Physics and Astronomy, The Johns Hopkins
  University, Baltimore MD 21218}

\paindex{Meurer, G.R.}

\authormark{G.R.\ Meurer}

\begin{abstract}
Observations with the ACS Wide Field Camera and G800L grism can produce
thousands of spectra within a single WFC field producing a potentially
rich treasure trove of information.  However, the data are complicated to
deal with.  Here we describe algorithms to find and characterize spectra
of emission line galaxies and supernovae using tools we have developed
in conjunction with off the shelf software.
\end{abstract}

%-----------------------------------------------------------------------
%                       Subject Index keywords
%-----------------------------------------------------------------------
% Enter up to 6 keywords describing your paper.  These will NOT be
% printed as part of your paper; however, they will be used to
% generate the subject index for the proceedings.  There is no
% standard list; however, you can consult the indices for past Calibration
% Workshop Proceedings. 

\keywords{ACS: grism, spectroscopy, aXe, star forming galaxies,
supernovae}

\section{Introduction}

The G800L grism combined with ACS's Wide Field Camera is a powerful
combination for obtaining thousands of spectra with relatively modest
outlay of HST time.  However, the resulting images are difficult to
interpret due to a number of peculiarities including: (1) strong
spatially varying sky background; (2) a position dependent wavelength
solution; (3) the wide spectral response: a three dimensional flatfield
and modeling of the wavelengths contributing to each pixel is required
for precise flatfielding; (4) tilted spectra with respect to the CCD
grid (the tilt varies over the field); (5) each source is dispersed into
multiple orders resulting in much overlap - deep images become confusion
limited; (6) zeroth order images of compact sources can easily mimic the
appearance of sharp emission features; and (7) the low resolution ($R
\approx 90$ for point sources) means that only high Equivalent Width
(EW) features can be discerned, while most familiar features are blends.
The Space Telescope European Coordinating Facility has done an excellent
job of addressing these peculiarities with the software package {\sl
aXe} (Pirzkal et al.\ 2001).  Armed with it and a broad band detection
image, users can extract 1D and 2D spectra that are sky-subtracted,
wavelength-calibrated, flatfielded, and flux calibrated with minimum
effort.  Here I describe complimentary techniques I have developed to
analyze WFC grism images. Specifically, I describe tools geared to
finding emission line sources and supernovae (SNe).  Here I concentrate
on my work with the ACS GTO team to search the Hubble Deep Field North
(HDFN) for Emission Line Galaxies (ELGs) and work with the PEARS team to
find SNe.

\section{Initial Processing}

\begin{figure}[tbp]
\epsscale{0.65}
\plotone{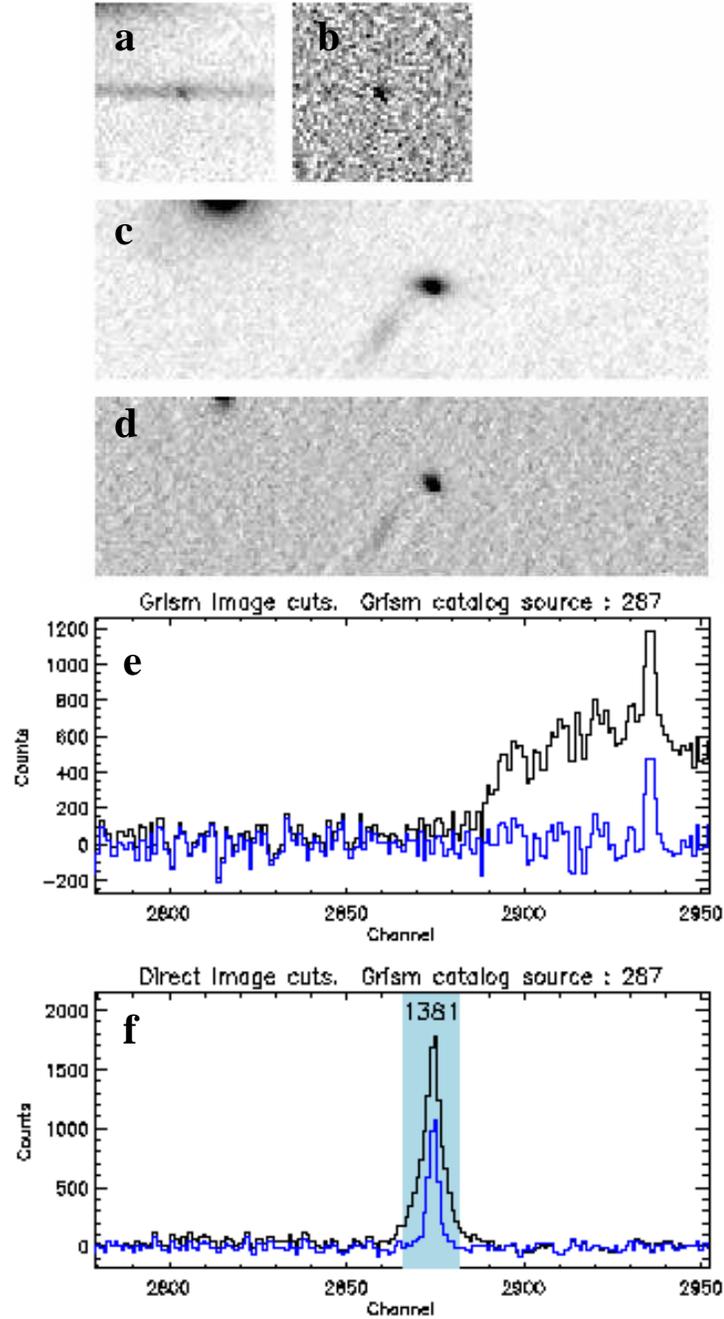}
\caption{Steps in processing the grism and broad band detection images
  for finding ELGs using method B. Panels a and b show a 50$\times$50
  cutout of the grism image before and after subtracting a 13$\times$3
  median filtered version of the image.  Panels c and d show cutouts of
  the detection image before and after the same filtering.  The width of
  the cutout covers the full $x$ range overwhich the counterpart to the
  source seen in panel b may reside. Panel e shows the collapsed 1D
  spectra of five rows centered on the emission line extracted from the
  grism image before (black line) and after (blue line) median
  filtering.  Panel f is the same but for the 1D cuts through the
  detection image.  The shaded region with identification is derived
  from collapsing the the {\sl SE\/} segmentation image of the detection
  image.}
\label{f:grexamp}
\end{figure}

{\sl aXe} is designed so that it can work with a stack of individual
dithered exposures (the FLT or CRJ images), where the grism images have
not been flatfielded nor geometrically corrected.  However, both
flatfielding and drizzling can be very useful.  Application of the F814W
flat corrects most gross blemishes and removes at least half the
amplitude of large scale sky variations (after geometric
correction).  Spurious dark spots may remain at the blue end of some
spectra, but their amplitude will be diluted if there are numerous small
dithers.  Their presence will have little impact on emission line
searches, while their sharpness means they are unlikely to be confused
with real absorption features. 

\clearpage

Drizzle combining multiple dithered exposures is feasible as long as the
dither offsets are all within 6$''$; then the alignment across the
spectra will all be correct to within 0.5 pixels.  The resultant
geometrically corrected images have first order spectra that are nearly
horizontal across the image, and greatly decreases the spatial variation
in the wavelength solution.  Drizzle combining also allows improved CR
rejection, especially when done with the ACS GTO pipeline {\sl Apsis\/}
(Blakeslee et al.\ 2002).

A mask is used to mark or remove the zeroth order images. First the
zeroth order sources in the grism image is matched to those in a broad
band detection image.  The sources are found with {\sl SExtractor\/}
(hereafter {\sl SE}; Bertin \&\ Arnouts, 1996) which is used to catalog
the sources in both the detection and grism images.  Only compact
sources are matched.  Their positions are used to define a linear
transformation between the detection image and the zeroth order.  The
scaling ratio between the matched detection and zeroth order images is
determined and used to model which pixels to mask.  In the HDFN the the
images in F775W and F850LP are typically 32 and 21 times brighter,
respectively, in count rate than the zeroth order counterparts.  This
scaling ratio is used to determine which pixels will have zeroth order
counterparts that are brighter than sky noise level.  The position of
these pixels in the detection image are transformed to populate a mask
for the grism image which is then grown by three pixels to account for
the slight dispersion in the zeroth order.  Masked pixels are set to
zero at the appropriate stage of the analysis.

% This pre-anlysis processing does not preclude the possibility of
% extracting spectra with {\sl aXe\/}; rather it eases its burden.
% Flatfielding is no longer required, and it does not need to process a
% stack of input spectra.  However when using {\sl aXe\/} the correct
% wavelength solution for geometrically corrected data must be employed.  

\section{Finding Emission Line Galaxies}

The ACS Science team observations centered on the HDFN consist of 3
orbits with G800L and F850LP ($z_{850}$) and two orbits with F775W
($i_{775}$).  Two complimentary techniques for finding ELGs were
employed on this field.

\parindent=0mm

{\bf A: Search 1D spectra}.  {\sl aXe\/} is used to extract spectra of
all {\sl SE\/} cataloged sources in the detection image down to $i_{775}
= 26.5$ ABmag.  The flux calibrated spectra are then filtered by
subtracting 13 pixel median smoothed spectra leaving only sharp
features.  Sources with peaks having $S/N > 4$ are flagged as likely
ELGs.  The flagged spectra are classified by eye - broad absorption line
sources are also flagged by this algorithm.  These are usually M or K
stars, but also include the two SNe in this field (Blakeslee et al.\ 2003) .
The true ELGs have their emission lines fitted with Gaussians to derive
line wavelength and flux.

\begin{figure}[tbp]
\epsscale{0.65}
\plotone{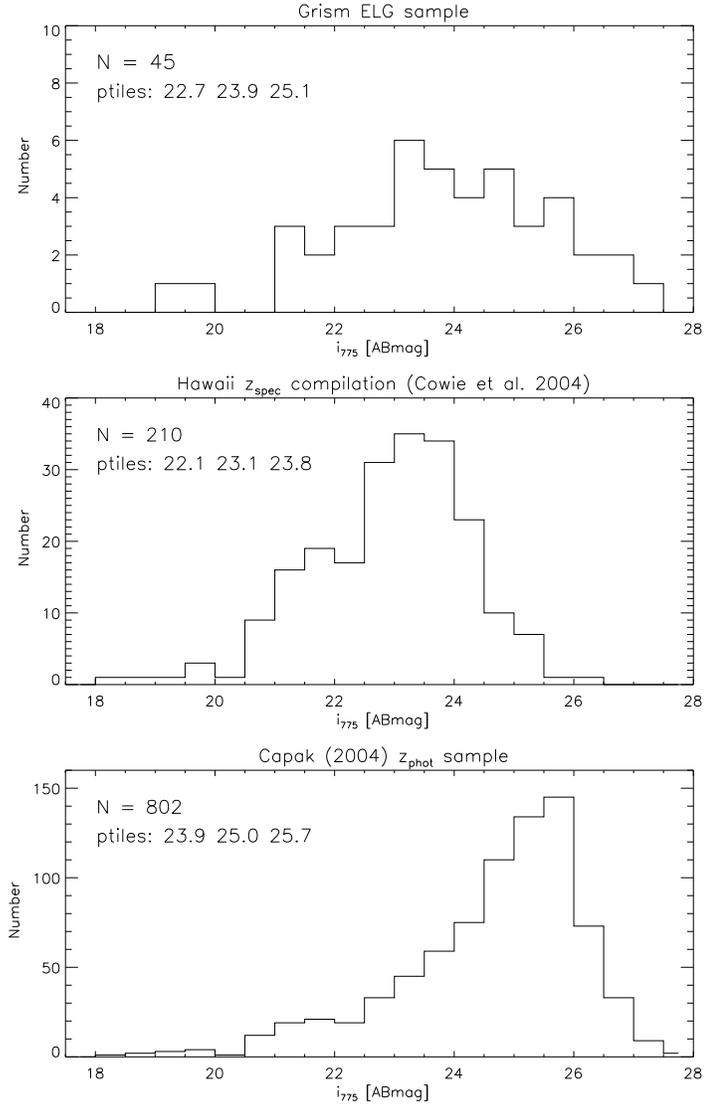}
\caption{Histogram of $i_{775}$ AB magnitudes of the grism selected ELG
  sample in the HDFN (top panel) compared with the spectroscopic
  redshift sample of Cowie et al.\ (2004; middle panel) and the
  photometric redshift sample of Capak (2004; bottom panel).  In the
  upper left of each corner we report the total number of sources in the
  sample and the 25th, 50th (median), and 75th percentile $i_{775}$
  AB magnitudes.}
\label{f:mhist}
\end{figure}

{\bf B: Search 2D grism image}.  The basis of this method is the
observation that most emission line sources appear to correspond to
compact knots, not necessarily at the center of galaxies.  Here we find
the line emission in the grism image first and then pinpoint the
emitting sources in the detection image, as illustrated in
Fig.~\ref{f:grexamp}.  Sharpened versions of both the grism and direct
images are made by subtracting a 13$\times$3 median smoothed version of
themselves.  In the grism image, this effectively subtracts the
continuum and removes cross-dispersion structure.  This image is then
cataloged with {\sl SE\/}.  Ribbons, typically covering five rows,
centered on the $y$ position of each source are extracted from both the
sharpened grism and direct images.  Since the dispersed spectrum lies
primarily to the right of the direct image, the extracted ribbons extend
more to the left so that all possible sources that could have created
the emission line are in the direct ribbon.  The ribbons are collapsed
down to 1D spectra and cross correlated after the regions beyond $\pm$
13 pixels from the source in the grism image are set to 0.0.  This is
done so they do not contribute to the cross-correlation amplitude.  Any
knot within the detection ribbon will produce a peak in the
cross-correlation spectrum.  The position of the peak yields the offset
between the knot and the line emission in the grism image.  Using the
wavelength solution for the grism, in principle one could derive the
line wavelength from this offset.  Instead, final measurements of the
emission line quantities are obtained from 1D spectra of each knot
extracted with {\sl aXe\/} using the cross-correlation determined
position of the star forming knot.  As with method A, the emission line
properties are measured with Gaussian fits to the spectra.

\parindent=2em

In the HDFN field we found 30:39 ELGs with methods A:B.  For the most
part, the same galaxies are found; 7:16 ELGs were
uniquely found with methods A:B.  For three ELGs we identified multiple
emission line knots with method B.  Figure~\ref{f:mhist} compares the
$i_{775}$ apparent magnitude distribution of the merged list of ELGs
from our analysis compared to galaxies with spectroscopic and
photometric redshifts in the same field.  The grism ELGs,
found with three orbits of HST time, are on average fainter than the
galaxies with spectroscopic redshifts gathered over several years from
the Keck 10m telescopes.

\section{Line Identification}

Line identification is a major concern.  Only seven of the ELGs in HDFN
have two emission lines in our data.  In those cases the lines can be
identified using the ratio of wavelengths which remains invariant with
redshift. However one must be careful with this technique since
$\lambda_{\rm H\alpha}/\lambda_{\rm [OIII]} = 1.3138$ is close to
$\lambda_{\rm H\beta}/\lambda_{\rm [OII]} = 1.3041$.  A one pixel
(42\AA) uncertainty in both line wavelengths could result in an
incorrect line identification.  

The remaining sources only have one line.  The dispersion is too low to
split the \fion{O}{II} doublet, the \fion{O}{III}4959,5007\AA\ lines are
also blended, as is \Halpha\ and the \fion{N}{II} doublet.  With only one
line, at the grism's resolution, then a good first guess redshift is
crucial for line identification.  Drozdovsky et al.\ (2005) tackle this
problem, in part, by looking at the size of the host galaxies.  However,
size alone is not a great indicator of redshift - there is little
evolution in angular size for $z > 0.2$.  Our approach is to use
photometric redshifts as the first guess redshift.  This results in 
line identifications for 37 of the 39 single line ELGs.

\begin{figure}[tbp]
\plottwo{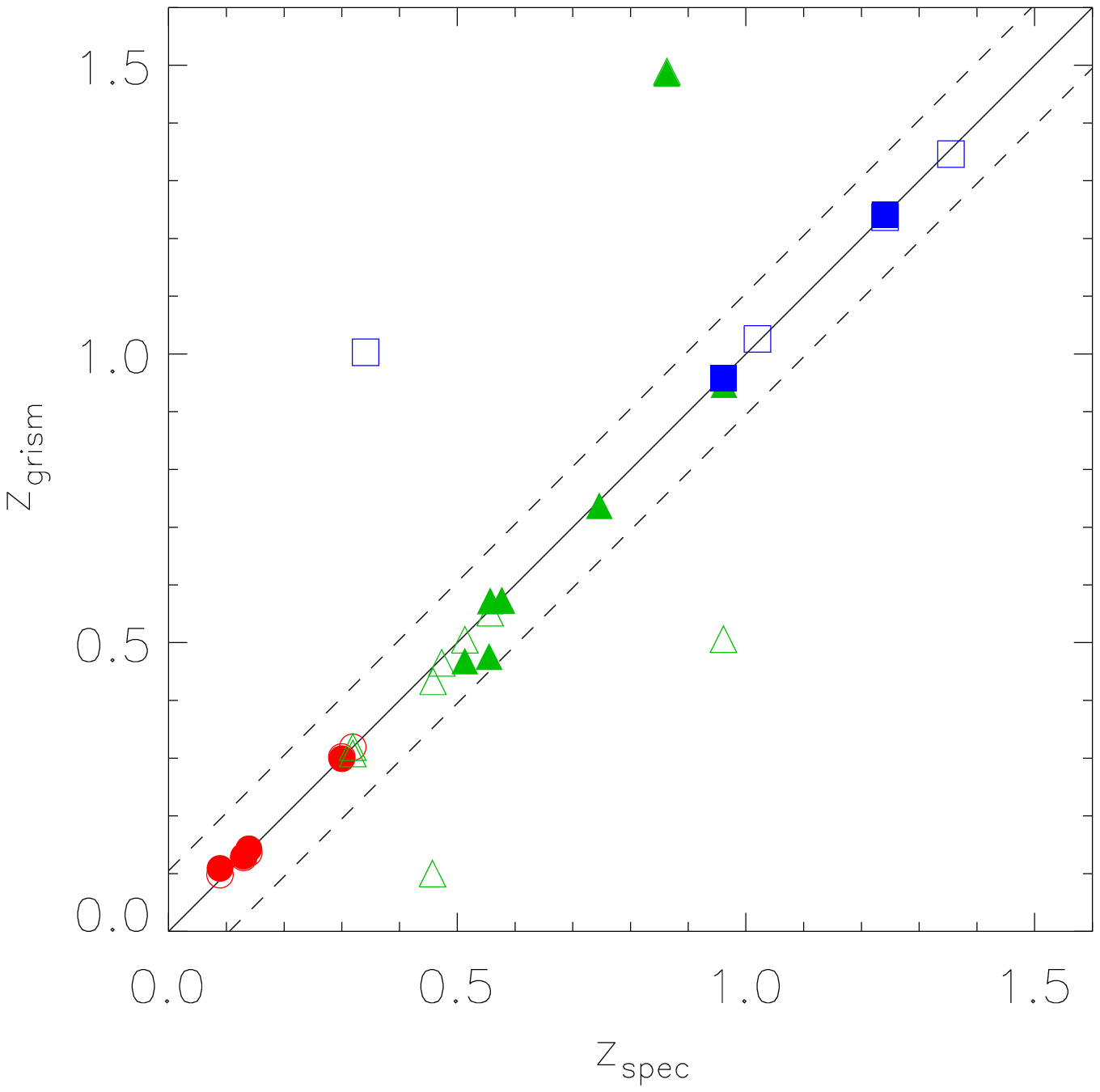}{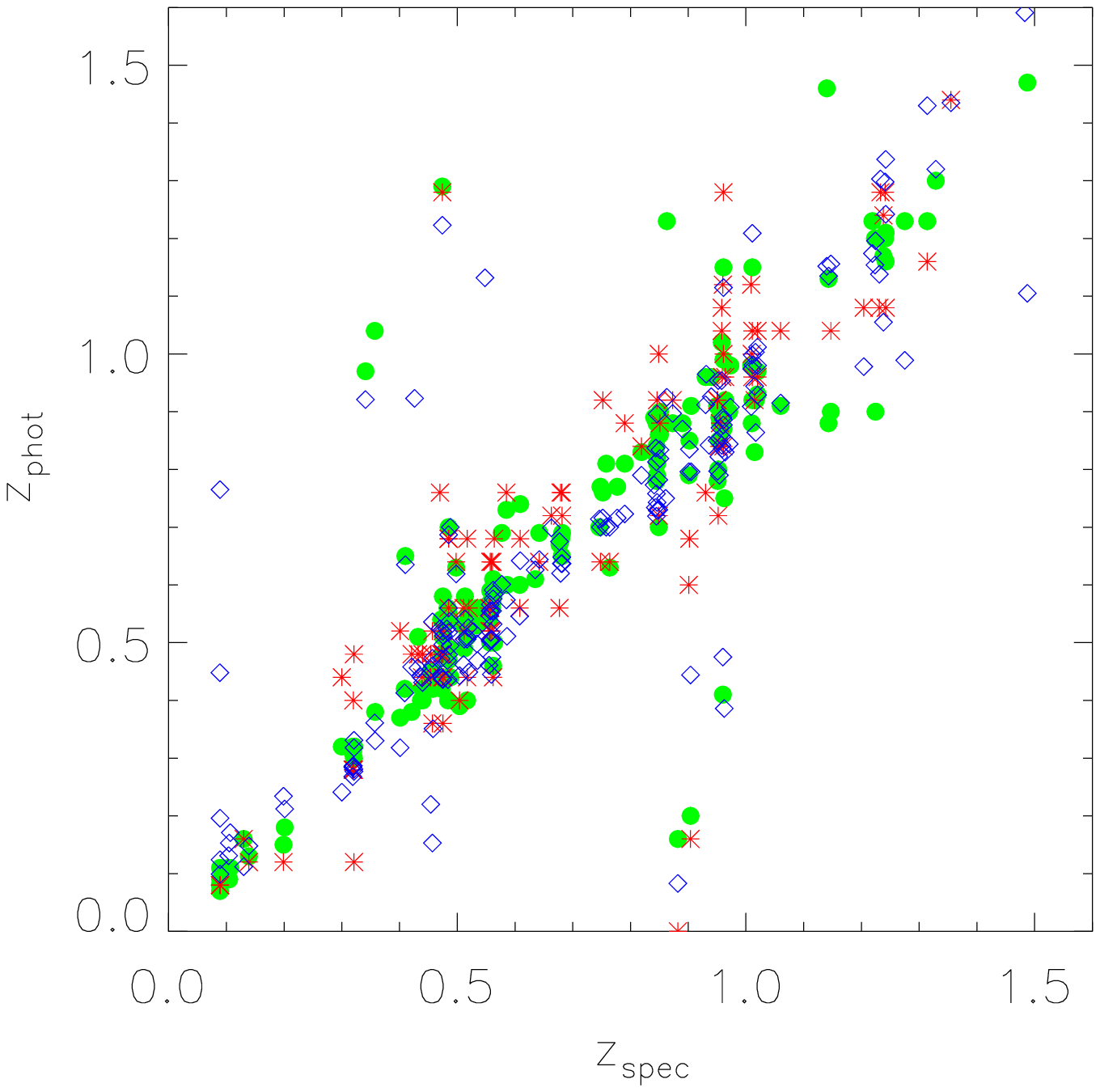}
\caption{Comparison of grism redshifts (right) and photometric redshifts
  (left) with spectroscopic redshifts from Cowie et al.\ (2004).  In the
  right panel, the unity relationship is shown as a solid line, sources
  outside the dashed lines at $\Delta z = \pm 0.105$ are outliers. Only
  photometric redshifts were used for the first guess redshift.  If
  spectroscopic redshifts are used as priors there is still one outlier.
  Here, measurements from method A are shown with solid symbols,
  meaurements from method B are shown as open symbols.  The symbol shape
  and color indicate the grism line identification: \Halpha\ emitters
  are (red) circles, \fion{O}{III} emitters are (green) triangles and
  \fion{O}{II} emitters are (blue) squares.  In the left panel the
  photometric redshifts from Cowie et al.\ (2004), Fernandez-Soto et
  al.\ (2004), and our measurements are shown as (green) filled circles,
  (red) asterisks, and (blue) hollow diamonds respectively.}
\label{f:zgr}
\end{figure}

Figure~\ref{f:zgr} compares grism redshifts with spectroscopic
redshifts, in panel a, and spectroscopic versus photometric redshifts in
panel b.  Taking the spectroscopic redshifts as ``truth'' results in
1/15 : 3/19 misidentified lines with methods A:B.  This is similar to
the error rates from photometric redshifts, as can be discerned from
Fig.~\ref{f:zgr}b.  The dispersion about the $z_{\rm grism}$ versus
$z_{\rm spec}$ unity line, excluding the outliers is 0.016:0.009 for
methods A:B.  Method B is probably more accurate because it better
pinpoints the location of line emission.  This compares to a dispersion
about the unity line in $z_{\rm phot}$ versus $z_{\rm spec}$ (after
clipping outliers) of 0.073, 0.107, 0.082 for $z_{\rm phot}$ estimates
from Capak (2004), Fernandez-Soto et al. (1999), and our own photometric
redshifts respectively.  Thus grism redshifts are nearly an order of magnitude
more accurate than photometric redshifts.  

\begin{figure}[btp]
\plotone{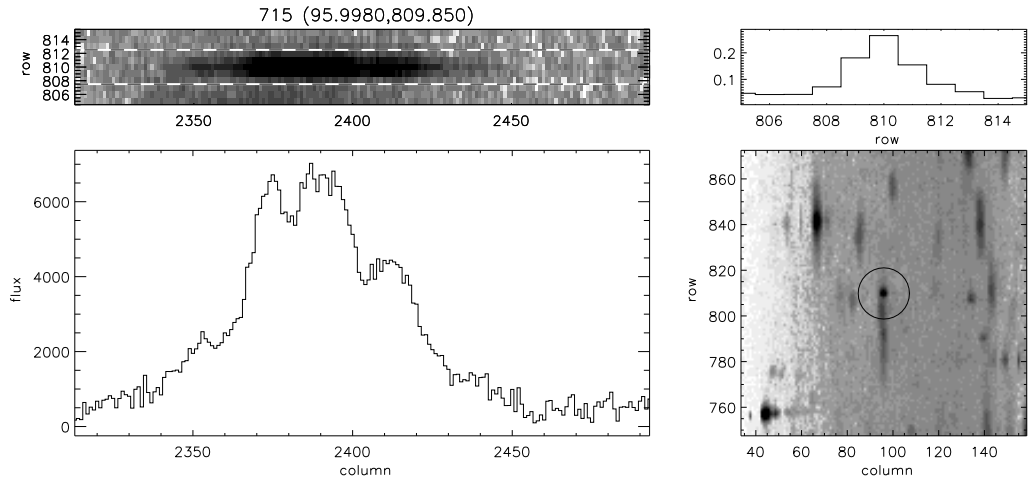}
\caption{Example of a supernova identified with {\sl SHUNT\/}.  The top
  left panel shows the geometrically corrected grism image.  The bottom
  left panel shows the extracted 1D spectrum found by collpasing the
  above 2D image between the dashed lines.  The top right panel shows a
  1D cut along the cross dispersion of the spectrum.  The bottom right
  panel shows the squashed grism image with the source identified.}
\label{f:snexamp}
\end{figure}

\section{Finding Supernovae}

The two SNe discovered in the HDFN have broad absorption features,
distinctly different from Galactic stars, and are easily visible in our
grism spectra (Blakeslee et al.\ 2003) demonstrating the viability of
grism surveys for SNe searches.  The PEARS team has obtained 200 orbits
of HST time primarily to characterize high-redshift objects in the two
GOODS fields using the WFC and G800L grism.  An additional aim is to
search for SNe on a rapid turn-around basis.  The total exposure time at
each pointing/roll angle is about twice as long as the HDFN observations
described above.  However, only shallow broad-band images are obtained
concurrently with the grism exposures.  These are used to align the
grism images to the astrometric grid of the GOODS fields.  But they are not as
deep as the grism image, hence they may not reveal SNe.  So although {\sl
aXe\/} spectra are generated of the prior GOODS cataloged sources, they
are not useful for finding SNe at later epochs.

What is needed is a method to find SNe using only the grism images.  To
this aim I have developed an IDL package {\sl SHUNT\/} (Supernovae Hunt)
to find and classify the first order spectra of all compact sources in a
grism field.  As before, the starting point is geometrically corrected,
combined grism images.  Since most source cataloging codes (i.e.\ {\sl
SE\/}) have been developed to find compact blobs, they do not work so
well for finding grism spectra which are very extended, often at near the
noise level of the image.  Rather than optimizing the code to fit the
data, {\sl SHUNT\/} makes the data suit the code by squashing (rebinning)
the image $25 \times 1$ per output pixel before cataloging with {\sl
SE\/}.  This results in first order spectra being close to critically
sampled in the $x$ direction.  The resultant catalog is filtered to
remove small sources (typically zeroth order images) and extended
sources (galaxies).  The remaining $\sim$ 250 sources are then
classified.  Five rows centered on each source are collapsed to form a
spectrum (which is not wavelength calibrated).  Classification is by eye
where the classifier (me) examines figures such as Fig~\ref{f:snexamp}
showing the 2D spectrum, the collapsed 1D spectrum, a cross dispersion
trace and the squashed grism image.  Each source is classified as either
SN, unidentified absorption spectrum, probable M or K star, break
spectrum, emission line source, featureless, non-first order spectrum,
or spurious (the order is of decreasing interest, and roughly of
increasing occurrence rate).  Direct postage stamp images from GOODS (or
the shallow broad-band images) are generated with a rectangular error
box plotted which should contain the source.  An empty error box in the
prior GOODS image is a second indication of a transitory source.  It
typically takes 0.5 to 1 hour to classify all the objects in a field.

\begin{figure}[tbp]
\plottwo{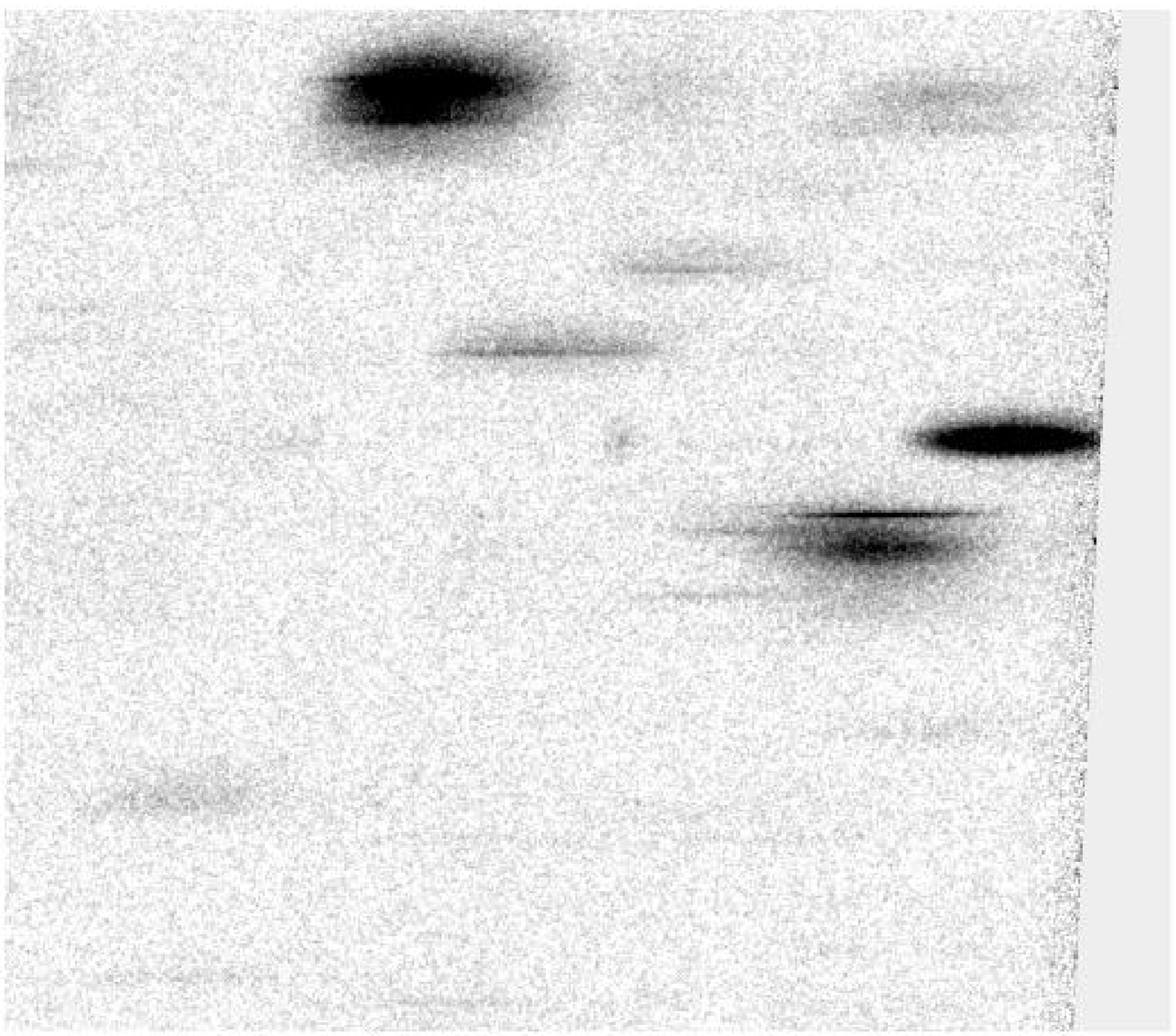}{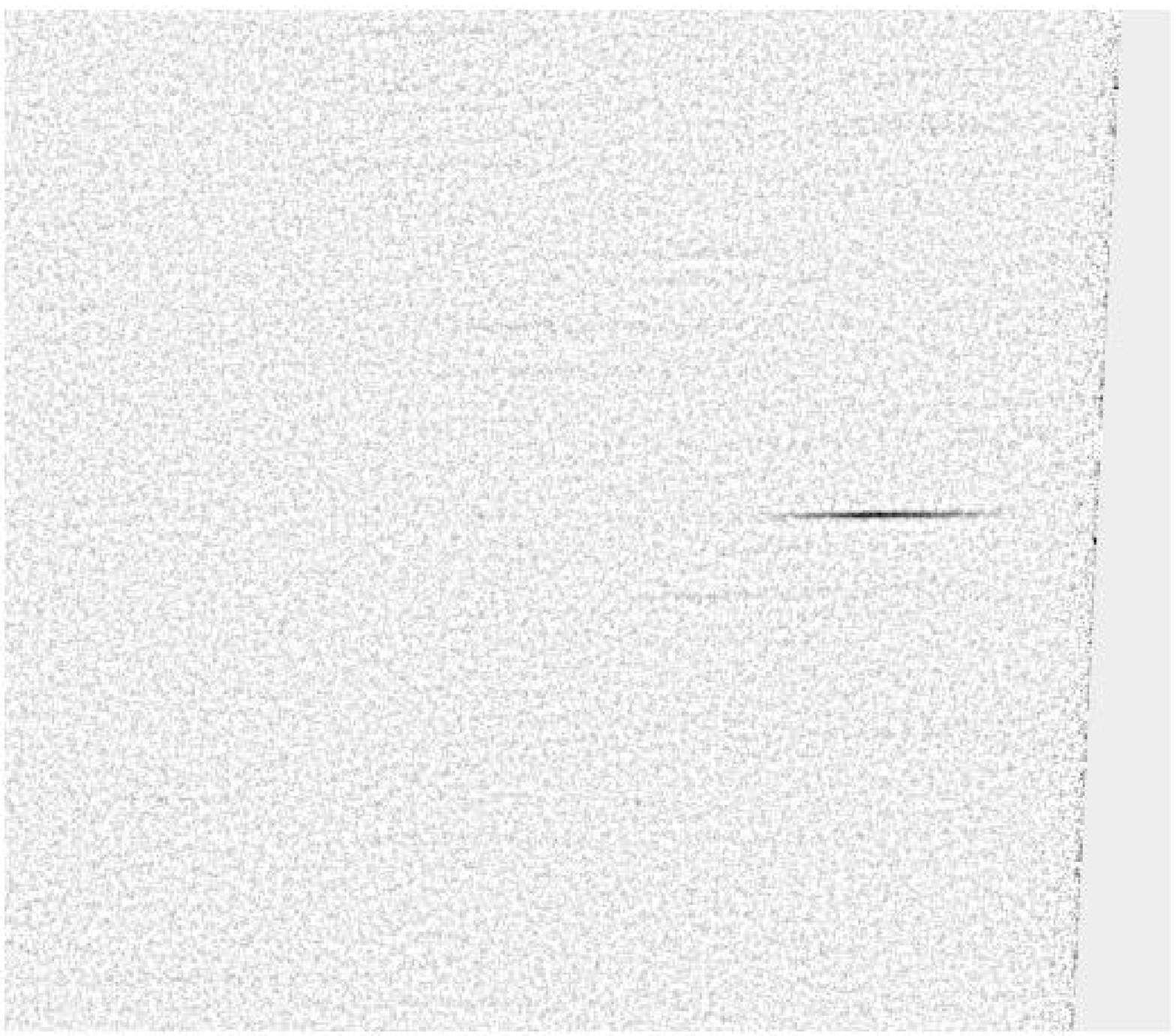}
\caption{Portion of a grism image containing a faint late type SN.  
  Panel a (left) shows the geometrically corrected and combined image.
  Panel b (right) shows the same image after subtracting the {\sl aXe\/}
  model image based on prior GOODS observations, and then subtracting a
  smoothed version of the residuals.  The SN now clearly stands out.}
\label{f:faintsn}
\end{figure}

One problem with this approach is that it can miss SNe blended with
galaxy spectra.  This is more likely to occur for late time SNe spectra
which can have low S/N and/or be featureless at grism resolution.  An
example is shown in Fig~\ref{f:faintsn}.  One way such objects can be
found is to subtract model spectra of the sources cataloged by GOODS
using {\sl aXe} v1.5 (K\"ummel, this volume).  The models are very good
but not perfect.  However, subtraction of the smoothed residuals is
sufficient to isolate faint transient object spectra from the model
residuals.

\section{Summary}

The ACS grism produces amazingly rich datasets.  While the data are
somewhat difficult to interpret, tools have been developed to make the
most use of these data.  Public access tools like {\sl aXe\/} are
readily available to remove most of these complications and extract 1D
and 2D spectra.  Here I have shown how some common manipulations of the
data (such as geometric correction and flatfielding) allow interesting
sources such as emission line galaxies and supernovae to be efficiently
found.

\acknowledgments\ Many members of the ACS and PEARS Science teams as
well as others have contributed to the work presented here.  I
particularly thank Zlatan Tsvetanov, Holland Ford, Caryl Gronwall, John
Blakeslee, Peter Capak, Sangeeta Malhotra, Norbert Pirzkal, Chun Xu,
Txitxo Benitez, James Rhoads, Jeremy Walsh, and Martin K\"ummel.

% That's the end of the main body of the paper.  Now we will have some
% back matter.


\begin{references}

\reference Blakeslee, J.P., Anderson, K.R., Meurer, G.R., Benitez, N.,
\&\ Magee, D. 2002, ASP Conf.~Ser.~295: ADASS XII, 257

\reference Blakeslee, J.P., et al. 2003, \apj, 589, 693

\reference Bertin, E.\ \&\ Arnouts, S. 1996, \aaps, 117, 393

\reference Capak, P.L. 2004, Ph.D. Thesis, U.\ Hawai`i, 

\reference Cowie, L.L., Barger, A.J., Hu, E.M., Capak, P., Songaila
A. 2004, \aj 127, 3137

\reference Drozdovsky, I., Yan, L, Chen, H.-W., Stern, D., Kennicutt, R.,
Spinrad, H., \&\ Dawson, S. 2005, \aj, 130, 1324

\reference Fern{\' a}ndez-Soto, A., Lanzetta, K.M., \&\ Yahil, A. 1999,
\apj, 513, 34

\reference Pirzkal, N., Pasquali, A., \&\ Demleitner, M. 2001, ST-ECF
Newsletter, 29, ``Extracting ACS Slitless Spectra with aXe'',
p.\ 5 ({\tt http://www.stecf.org/instruments/acs}) 

\end{references}
\end{document}